\documentclass[twocolumn,apj]{emulateapj}

\usepackage{hyperref}

\usepackage{amsmath}

\begin{document}

\title{NuSTAR detection of X-ray heating events in the quiet Sun}

\author{Matej Kuhar\altaffilmark{1,2}, S\"am Krucker\altaffilmark{1,3}, Lindsay Glesener\altaffilmark{4}, Iain G. Hannah\altaffilmark{5}, Brian W. Grefenstette\altaffilmark{6}, David M. Smith\altaffilmark{7}, Hugh S. Hudson\altaffilmark{3, 5}, Stephen M. White\altaffilmark{8}}
\altaffiltext{1}{University of Applied Sciences and Arts Northwestern Switzerland, Bahnhofstrasse 6, 5210 Windisch, Switzerland}
\altaffiltext{2}{Institute for Particle Physics and Astrophysics, ETH Z\"{u}rich, 8093 Z\"{u}rich, Switzerland}
\altaffiltext{3}{Space Sciences Laboratory, University of California, Berkeley, CA 94720-7450, USA}
\altaffiltext{4}{School of Physics and Astronomy, University of Minnesota - Twin Cities , Minneapolis, MN 55455, USA}
\altaffiltext{5}{SUPA School of Physics \& Astronomy, University of Glasgow, Glasgow G12 8QQ, UK}
\altaffiltext{6}{Cahill Center for Astrophysics, 1216 E. California Blvd, California Institute of Technology, Pasadena, CA 91125, USA}
\altaffiltext{7}{Physics Department and Santa Cruz Institute for Particle Physics, University of California, Santa Cruz, 1156 High Street, Santa Cruz, CA 95064, USA}
\altaffiltext{8}{Air Force Research Laboratory, Albuquerque, NM, USA}

\begin{abstract}
The explanation of the coronal heating problem potentially lies in the existence of nanoflares, numerous small-scale heating events occuring across the whole solar disk. In this paper, we present the first imaging spectroscopy X-ray observations of three quiet Sun flares during the NuSTAR solar campaigns on 2016 July 26 and 2017 March 21, concurrent with SDO/AIA observations. Two of the three events showed time lags of a few minutes between peak X-ray and extreme ultraviolet (EUV) emissions. Isothermal fits with rather low temperatures in the range $3.2-4.1$ MK and emission measures of $(0.6-15)\times10^{44} \textrm{ cm}^{-3}$ describe their spectra well, resulting in thermal energies in the range $(2-6)\times10^{26}\textrm{ ergs}$. NuSTAR spectra did not show any signs of a nonthermal or higher temperature component. However, since the estimated upper limits of (hidden) nonthermal energy are comparable to the thermal energy estimates, the lack of a nonthermal component in the observed spectra is not a constraining result. The estimated GOES classes from the fitted values of temperature and emission measure fall between $1/1000 \textrm{ and } 1/100$ A class level, making them 8 orders of magnitude fainter in soft X-ray flux than the largest solar flares. 
 \end{abstract}
 

\section{Introduction}
The explanation of how the corona keeps its temperature of a few million Kelvin, termed the `coronal heating problem', has eluded scientists for decades. Since solar flares release energy and heat ambient plasma, it is argued that they may provide (at least a part of) the needed energy to sustain coronal temperatures. 
\par Solar flares follow a negative power-law frequency distribution with increasing energy, with a power-law index $\sim2$ \citep[e.g., ][]{Hudson91, Hannah08_2}. A flat distribution, with a power-law index below 2, implies that smaller events do not dominate the energy released in flares. Since the largest flares do not occur frequently enough to heat the solar corona, it has been instead argued that smaller-scale reconnection events could have a steeper frequency distribution, providing the needed energy input due to large numbers. \cite{Parker88} introduces the term \textit{nanoflares} for such events, with energies speculated to be of the order of $10^{24} \textrm{ ergs}$ or less, as estimated from ultraviolet fluctuations within active regions \citep{Porter84}. This triggered many theoretical studies on the role of small-scale events in coronal heating \citep[e.g., ][]{Walsh03, Klimchuk06, Browning08, Tajfirouze12, Guerreiro15, Guerreiro17}.

\par Parker's basic magnetic energy releases, however, are yet to be confirmed observationally, most probably due to their modest sizes and energies, combined with sensitivity limitations of present solar instruments. So far, only measurements of individual events down to $\sim 10^{24}$  ergs (at the `high-energy' end of Parker's estimate) have been performed, while less energetic nanoflares could have even smaller energies and should form an ensemble of indistinguishable reconnection and heating processes that make the solar corona. In addition to searches for nanoflares in soft X-rays \citep[e.g., ][]{Shimizu97, Katsukawa01, Terzo11}, the most complete statistical study of microflares in hard X-rays is by \cite{Hannah08_2}, using 6 years of \textit{Reuven Ramaty High Energy Solar Spectroscopic Imager} \citep[RHESSI, ][]{Lin02} data and including more than $25 \hspace{3pt} 000$ microflares. However, since RHESSI is sensitive to flares with temperatures above $\sim10$ MK and emission measures (EM) above $10^{45}\textrm{ cm}^{-3}$, the events included in the above study are much larger and more energetic than nanoflares proposed by \cite{Parker88}. Another distinctive feature is that RHESSI observes microflares only from active regions, while nanoflares should occupy the whole solar disk. Quiet Sun (QS) flares, on the other hand, have been observed only in soft X-rays (SXR) and extreme ultraviolet (EUV) narrow-band filter observations \citep[e.g., ][]{Krucker97, Krucker98, Parnell00, Aschwanden00}. These brightenings have been found to occur on the magnetic network of the QS corroborating the magnetic energy releases as their drivers. Radio events in the GHz range associated with the EUV brightenings have been speculated to be signatures of non-thermal electrons accelerated during the energy release process \citep{Benz99}. Their spectroscopic X-ray signatures, however, are  too faint for the state-of-the-art solar X-ray instruments. Therefore, in order to confirm Parker's nanoflare scenario of coronal heating, it is crucial to perform sensitive imaging spectroscopy X-ray observations of small-scale events across the whole solar disk. 
\begin{figure}[hbtp]
\centering
\includegraphics[scale=0.41]{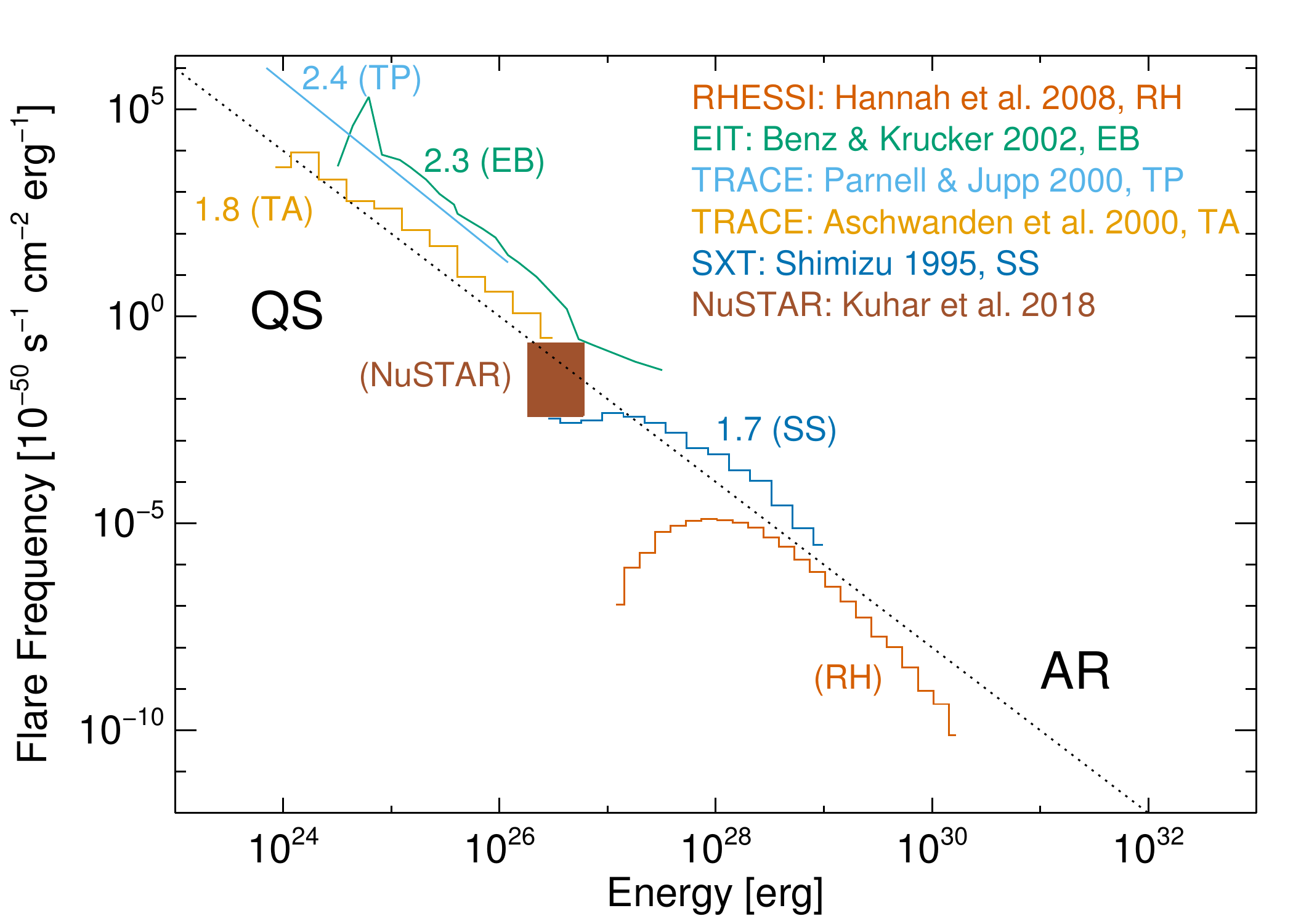}
\caption{Flare frequency distribution vs. energy from various X-ray and EUV studies. NuSTAR observations analyzed in this paper are presented as the brown rectangle. Note that the presented studies used data from different phases of the solar cycle, making comparisons of the flare occurence between them difficult. The dotted line shows one frequency distribution with a power-law index of 2 to guide the eye. Taken from \cite{Hannah11} and adapted to include our results.}
\end{figure}

\begin{figure*}[hbtp]
\centering
\includegraphics[scale=0.88, trim=200 80 200 80]{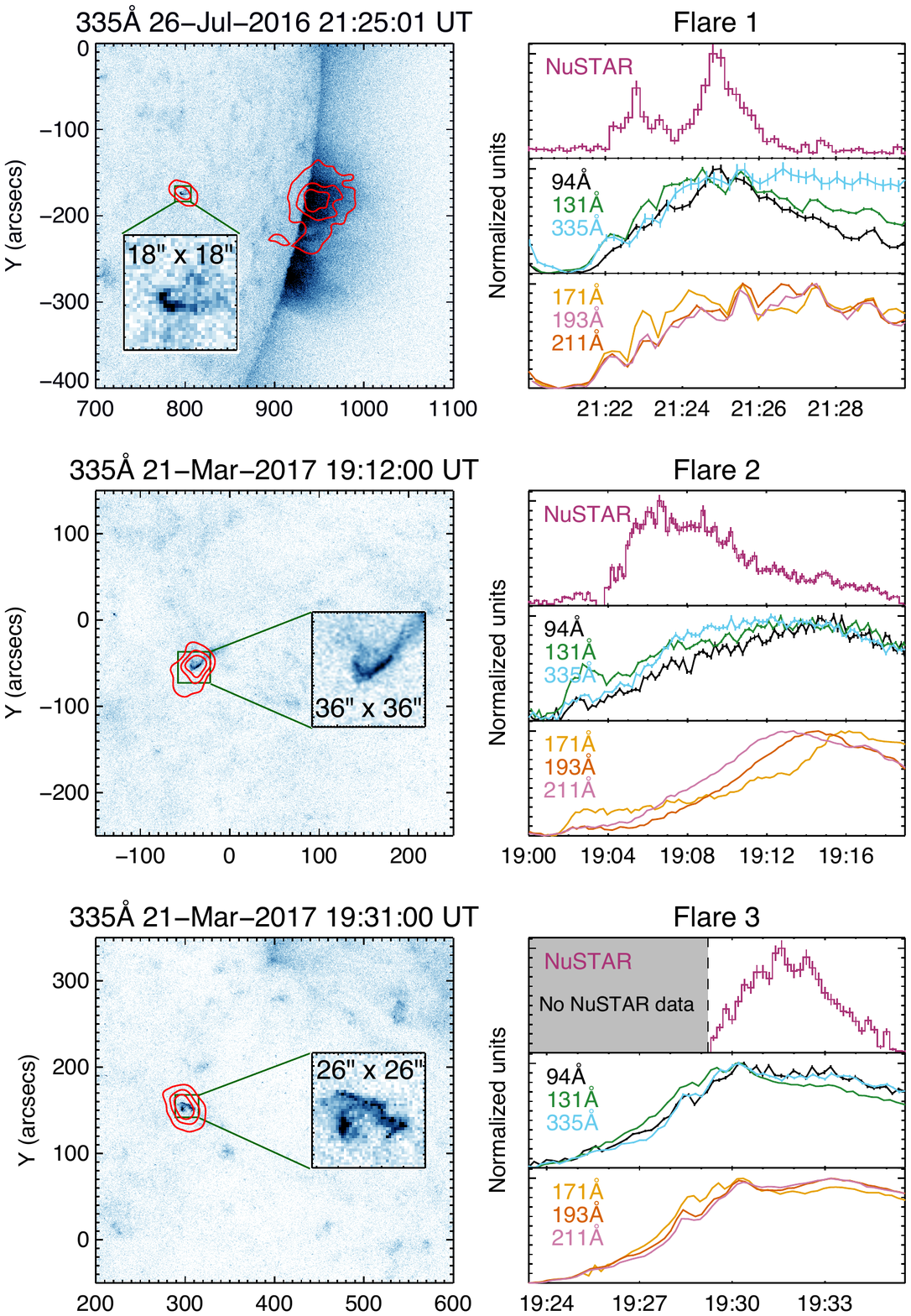}
\caption{Overview plots of the three QS flares. \textit{Left panels:} $400''\times400''$ AIA $335 \textrm{\AA}$ images of the events, together with zoomed-in images of the event morphology in the insets. The 30, 50 and 70\% contours of maximum NuSTAR emission are shown in red. \textit{Right panels:} Background-subtracted time evolution of the flaring region and flux uncertainties in the combined flux of NuSTAR focal plane modules A and B above 2.0 keV together with AIA $94 \textrm{\AA}$, $131 \textrm{\AA}$, $335 \textrm{\AA}$, $171 \textrm{\AA}$, $193 \textrm{\AA}$ and $211 \textrm{\AA}$ channels. Error bars in 171$\textrm{\AA}$, 193$\textrm{\AA}$ and 211$\textrm{\AA}$ channels are smaller than the line thickness.}
\end{figure*}

\par \textit{The Nuclear Spectroscopic Telescope ARay} (NuSTAR)  is a focusing optics hard X-ray telescope launched in 2012 and operating in the energy range $3-79$ keV \citep{Harrison13}. Even though not solar-dedicated, it is capable of observing the Sun \citep[][]{Grefenstette16}, providing much higher sensitivity compared to indirect imaging telescopes such as RHESSI. It can therefore bridge the gap towards imaging spectroscopy in X-rays of small-scale heating events in the QS, and provide the opportunity to search for nonthermal signatures in them. This can be seen in Figure 1, where we show flare frequency distributions from various X-ray and EUV studies of microflares and QS brightenings \citep[][]{Shimizu95, Aschwanden00, Parnell00, Benz02, Hannah08_2}. The plot can be divided in two segments, the left one showing EUV observations of flares in the QS and the right one showing X-ray observations of microflares from active regions. QS NuSTAR observations from this study are shown by the brown box.
\par In this letter, we present first spectroscopically resolved X-ray measurements of QS flares. NuSTAR observations of QS heating events are described in Section 2. Data analysis and spectral fitting of the events is found in Section 3, while the discussion on this and possible future studies is presented in Section 4.

\section{Observations}
The data analyzed in this article were obtained in NuSTAR solar campaigns carried out on 2016 July 26 and 2017 March 21\footnote{Extensive information about all NuSTAR solar campaigns can be found at: \url{http://ianan.github.io/nsigh_all/.}}. Three QS events were observed during 1.5 hours of analyzed NuSTAR observations, one on 26 July 2016 and two others on 21 March 2017. They will be referred to as flares 1, 2 and 3 in the future sections, based on their chronological order. 
\par Figure 2 shows the spatial structure and time evolution for each of the events. Left panels show \textit{Atmospheric Imaging Assembly} \citep[AIA,][]{Lemen12} $335 \textrm{\AA}$ images of the part of the solar disk where the events occurred, together with the 30\%, 50\% and 70\% NuSTAR contours in red. NuSTAR images have been shifted to match the flare locations in AIA images in order to accommodate for uncertainties in absolute pointing \citep{Grefenstette16}. A zoomed-in image of each event is shown in the inset. Right panels show the time evolution of NuSTAR flux above 2.0 keV, as well as the time evolution of AIA EUV channels. All fluxes are background-subtracted, where background is defined as the lowest emission time frame during the pre-event phase.
\subsection{Time evolution}
Time profiles of flares generally reveal different behaviors for the thermal and nonthermal X-ray component. Non-thermal emissions are most prominently observed during the rise phase of the thermal emission (‘impulsive phase’) and can show several peaks with durations from a minute down to subsecond time scale \citep[e.g., ][]{Aschwanden95}. The main thermal emission evolves more gradually with a time profile often similar to the integrated nonthermal flux \citep[`Neupert effect', ][]{Neupert68}. (Hard) X-ray peaks that occur before the thermal peak (seen in soft-X-rays and/or EUV) are therefore often interpreted as a signature of nonthermal emission \citep{Veronig05}, but such a classification is not conclusive. Time lags between X-ray and EUV emission can also be produced by the different temperature sensitivity of X-ray and EUV observations: the X-ray peak is produced by the flare-heated plasma, which then cools to lower temperatures visible in EUV. To resolve the ambiguities present in the time evolution of X-ray and EUV emission, a spectral analysis is required. In the following, we discuss the time evolution of the individual events focusing on potential nonthermal signatures, followed by the spectral analysis in Section 3.
\par Flare 1 shows an intriguing time evolution with two distinctive X-ray peaks, while flares 2 and 3 have one broad peak dominating both the X-ray and EUV evolution. Flare 3 shows simultaneous X-ray and EUV peaks, in contrast to flares 1 and 2 which show a time lag of a few minutes between peak X-ray and EUV emissions. The rise of the EUV emission, as well as the decay, is slower than in X-rays for all flares. In order to interpret the observed relative timing, it is important to consider the difference in temperature responses between NuSTAR and AIA. NuSTAR has a steeply increasing response towards higher temperatures between 1 and 10 MK, making it sensitive primarily to the highest temperature plasma in this range. The AIA temperature response, on the other hand, is much broader and the resulting flux represents contributions from plasma at various temperatures. The time evolution of flare 2 can be explained by the process of plasma cooling, where NuSTAR peaks first, followed by the AIA channels according to their temperature sensitivity. The other events are more complex, and only a detailed temporal and spatial differential emission measure analysis might allow us to understand their complicated time evolution, but this is outside the scope of this letter. The spectral analysis presented in Section 3 further addreses the question of whether the delays between NuSTAR and AIA peaks imply nonthermal emission in these events.

\subsection{Flare locations and morphology}
Flare locations and morphologies can be found in the insets of left panels in Figure 2. Flare 1 evidenced an ejection of material during the impulsive phase, seen in all AIA channels. It occurred in the quiet Sun. Flare 2 was a part of a long lasting, elongated structure located in proximity to the solar disk center, with the flaring area just a fraction of the whole structure. The morphology of the structure is reminiscent of heated flare loops. Flare 3 was a short duration event that, like Flare 1, was not associated with any kind of X-ray or EUV structure. However, it showed an even more complex structure than flare 1. The March events were clearly associated with the quiet Sun magnetic network structures, while the association is not as clear for the July event. However, this might be due to its proximity to the solar disk, where the line-of-sight effects could mask the signal.
\par To conclude, in spite of their modest sizes and emission, the observed events show very complex spatial and temporal morphologies and therefore cannot be described as ``elementary'' energy releases proposed by Parker. They were not part of active regions and are therefore classified as QS events.
\section{Data Analysis}
\subsection{Spectra}
NuSTAR allows us to produce spectra for any time range, energy range (above 2.5 keV) and area. For our study, we use circular regions with diameter 55'' (a value close to NuSTAR's half power diameter) at each flare's location. Integration times were chosen individually for each flare so that the majority of X-ray emission is included (presented spectra are flare-integrated) and are equal to 4, 8 and 3 minutes for flares 1, 2 and 3, respectively. To perform spectral fitting in XSPEC \citep{Arnaud96}, NuSTAR spectra and response matrix files were obtained using standard NuSTAR data analysis software\footnote{https://heasarc.gsfc.nasa.gov/docs/nustar/analysis/}. In the following, we perform simultaneous fitting in XSPEC on the data from both focal plane modules, which are then combined to display the results shown in Figure 3 and Table 1. We fit an isothermal \citep[APEC in the XSPEC package, using abundances from][]{Feldman92} plus a fixed background model between 2.5 and 5.0 keV, where we estimate the background as a 2-minute integrated emission in the pre-flare phase, mostly consisted of ghost-rays (photons from sources outside the field-of-view).

\par NuSTAR spectra are shown in Figure 3. The fits give temperatures of $3.96^{+0.05}_{-0.40}$,  $4.01^{+0.05}_{-0.22}$ and $3.28^{+0.13}_{-0.06}$ MK, while their EMs lie in the range $5.6\times10^{43}-1.5\times10^{45} \textrm{cm}^{-3}$. These values of temperature and emission measure place our events just in between the active region microflares and the quiet Sun events analyzed previously in the EUV. Our events are at or slightly below the NuSTAR detection limit as derived from previous observations with lower livetime and much stronger ghost-ray signal \citep{Marsh17}. Here we note that the estimated EM for flare 2 is probably a lower limit, since we estimate that up to 50\% of the total flare emission might not be accounted for in our fits. This is due both to its proximity to the chip gap and a lot of changes in the combination of NuSTAR camera head units (CHU) used for pointing, which resulted in many (abrupt) changes in the estimated flare location. This has probably no effect on the temperature estimates, but the actual EM is likely a factor of 2 larger than the one reported. This is also shown in Table 1, with a factor 2 in parenthesis for parameters affected by this effect. The above reported temperatures and EMs place the observed events in the estimated range between 1/1000 and 1/100 GOES A-class equivalents, or between 7 and 8 classes fainter than the largest solar flares. 
\par It is interesting to note the low temperatures of NuSTAR QS flares. While RHESSI is designed to observe flares with temperatures above 10 MK, NuSTAR is able to observe lower temperatures due to its higher low-energy sensitivity. However, since NuSTAR's sensitivity also increases with increasing temperature, the fit-determined temperatures are the highest temperatures (as weighted by emission measure) present in the events. Therefore, it seems that QS flares reach only modest temperatures compared to those generally observed in regular active region flares. The only other possibility is that hotter QS events have significantly lower EMs, making them hard to observe even with NuSTAR. 

\begin{figure*}[hbtp]
\centering
\includegraphics[scale=0.92]{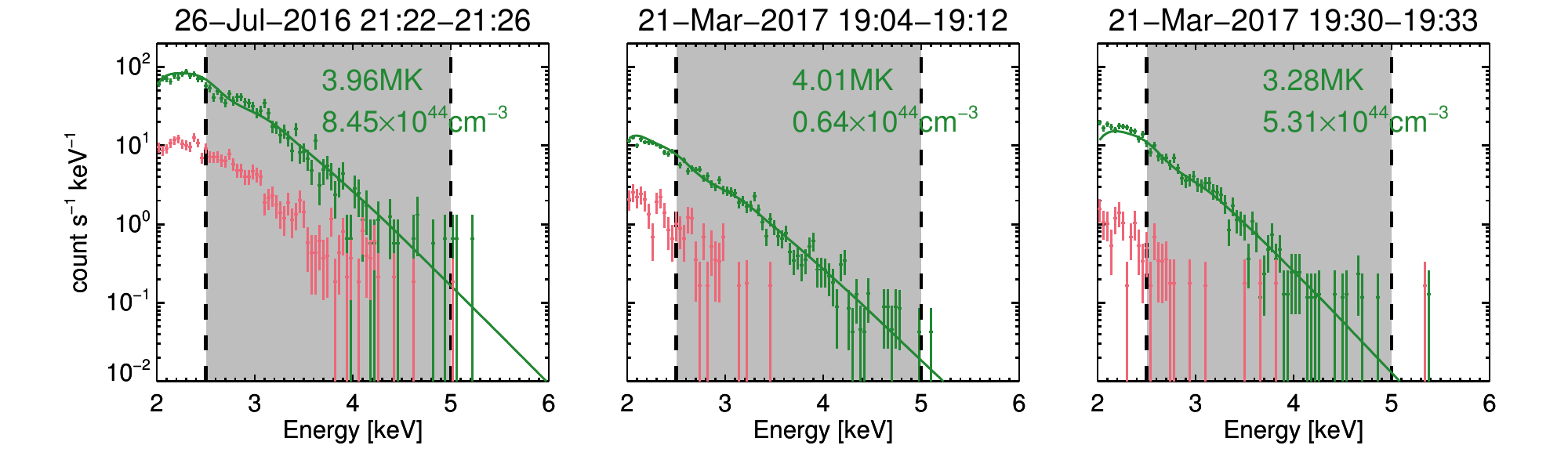}
\caption{NuSTAR spectra of the observed QS flares. Spectra with best isothermal fits for NuSTAR focal plane modules A and B combined is shown in dark green, while the background counts are shown in pink. The energy range $2.5-5.0$ keV used for spectral fitting is denoted by the grey area between the vertical dashed lines.}
\end{figure*}

\subsection{Thermal energy content}
\par We use the standard approach of estimating total thermal flare energy content from the flare plasma at the highest temperature as derived from NuSTAR spectra. This approach assumes that any cooler plasma such as observed in the EUV is a result of the cooling process. \cite{Wright17} estimated that this approximation could be up to $\sim30$\% different from the estimate from a complete, differential emission measure analysis of multithermal plasma in an active region microflare observed with NuSTAR and AIA. In this approximation, the thermal energy content of an event with temperature \textit{T}, emission measure \textit{EM} and volume \textit{V} is given by the formula \citep[e.g., ][]{Hannah08_2}
\begin{equation}
E_{th}\sim3NkT=3kT\sqrt{EM \cdot V.}
\end{equation}
To estimate upper and lower limits on the total thermal energy content, we use the combination of maximum and minimum of possible values for temperature and EM as given by the fits.  
\par Since the observed QS flares are not spatially resolved with NuSTAR, we estimate flare volumes as the area of flaring 335$\textrm{\AA}$ pixels (other channels have similar flaring areas) to the power of 3/2. As NuSTAR is only sensitive to the hottest plasma while AIA is sensitive to a broader range of temperatures, this estimate provides an upper limit for the actual volume and, consequentially, a lower limit for the density and an upper limit for the thermal energy content (an overestimate up to a factor of 5 in the thermal energy content is possible). Density estimates can be calculated with the formula $n=\sqrt{EM/V}$ and fall in the range $(0.5-4)\times10^9\textrm{ cm}^{-3}$. These values are similar to those derived from SXR QS flares by \cite{Krucker97} ($(1-5)\times10^{9}\textrm{ cm}^{-3}$), but larger than densities derived from EUV QS events by \cite{Aschwanden00} ($(0.1-0.5)\times10^{9}\textrm{ cm}^{-3}$). We calculate the following thermal energy contents for flares 1, 2 and 3: $(3.8-6.0) \times 10^{26}$, $(1.8-2.5) \times 10^{26}$ and $(3.9-5.9) \times 10^{26} \textrm{ ergs}$. These values are about 5 orders-of-magnitude smaller than in largest solar flares.

\subsection{Nonthermal emission}
There is no evidence for a high temperature or a nonthermal component in the spectra presented in Figure 3, and no counts above $\sim5$ keV are observed. By setting an upper limit for the potentially hidden nonthermal contribution, we estimate an upper limit of the energy in nonthermal electrons in the same way as has been in done in \cite{Wright17} and taking flare 1 as an example. The addition of a hidden nonthermal component with a low-energy cutoff at 5 keV and a power-law index of 7 still reproduces the observed spectrum well, giving undetectable signal above the cutoff. The estimated upper limit of the nonthermal energy equals $\sim5\times10^{26} \textrm{ ergs}$, a value within the uncertainties of the estimated thermal energy. Hence, the non-detection of a nonthermal component in the observed spectra is not a constraining result, with its upper limits still consistent with the observed heating.
\begin{center}
\begin{table*}
\centering
\label{table1}
\begin{tabular}{|c|c|c|c|c|c|c|c|c|c|}
\multicolumn{10}{c}{} \\
\hline   
  Flare &   Date    & Time & Location & Area &  Temperature &Emission measure &Density  &Energy  & GOES class    \\
   & [yyyy/mm/dd]      & [hh:mm]  & [x, y] & [arcsec$^2$]&  [MK] &[$10^{44}\textrm{ cm}^{-3}]$ &$[10^9 \textrm{ cm}^{-3}]$   &$10^{26}$ [erg]  &  [A]   \\
\hline 
   & & & & & &&&&\\
 1 & 2016/07/26 & 21:24&[795, -175] &38& $3.96^{+0.05}_{-0.40}$&$8.5^{+6.3}_{-0.9}$& $3.0^{+1.0}_{-0.2}$ & $4.5^{+1.5}_{-0.7}$& 0.01\\
    & & & & & &&&&\\

 2 & 2017/03/21 & 19:04&[-40, -55] &75& $4.02^{+0.05}_{-0.22}$&$(2\times)\textrm{ }0.64^{+0.22}_{-0.08}$& $(\sqrt{2}\times)\textrm{ } 0.51^{+0.08}_{-0.03}$ &$(\sqrt{2}\times)\textrm{ } 2.1^{+0.4}_{-0.2}$& $(2\times)\textrm{ } 0.0009$\\
    & & & & & &&&&\\

3 & 2017/03/21 & 19:30&[300, 150] &85& $3.28^{+0.13}_{-0.06}$&$5.3^{+1.8}_{-1.8}$& $1.3^{+0.2}_{-0.3}$ &$5.4^{+1.1}_{-1.1}$& 0.003\\
   & & & & & &&&&\\
\hline
\end{tabular}
\caption{QS flare parameters.}
\end{table*}
\end{center}
\section{Discussion and conclusions}
In this letter, we analyzed three QS flares observed in X-rays above 2.0 keV with NuSTAR. We were able to measure their X-ray spectra for the first time and derive flare peak temperatures (see Table 1 for the summary of the derived parameters). Despite their modest sizes and X-ray emission, these events show very complex spatial morphologies in the EUV. They are therefore not elementary energy releases and still much larger than Parker's idea of nanoflares.

\par Figure 4 shows our events in the $T-EM$ parameter space, together with two NuSTAR active region microflares observed in previous campaigns \citep{Glesener17, Wright17}. The green box represents SXR QS events from \cite{Krucker97}, showing they reach even lower temperatures and are below the sensitivity limits of our current observations.
 The isocurves show GOES classes, while the yellow area denotes the parameter space observable by RHESSI. For flares with temperatures between 3 and 4 MK as discussed here, RHESSI is sensitive to  EMs above $\sim10^{49}\textrm{ cm}^{-3}$, meaning  we gained at least four orders-of-magnitude in EM sensitivity compared to RHESSI. Another interesting result are rather low temperatures of up to $\sim4$ MK, indicating that QS flares might be reaching lower temperatures than the ones generally observed in active region flares.
\par In contrast to hints coming from the time evolution of NuSTAR and AIA fluxes, NuSTAR spectra did not show any sign of a high-temperature or a nonthermal component. However, since the estimated upper limits of energy in the hidden nonthermal component are comparable to the calculated thermal energies, the lack of a nonthermal component is not a strong diagnostic result.

\begin{figure}[hbtp]
\centering
\includegraphics[scale=0.58]{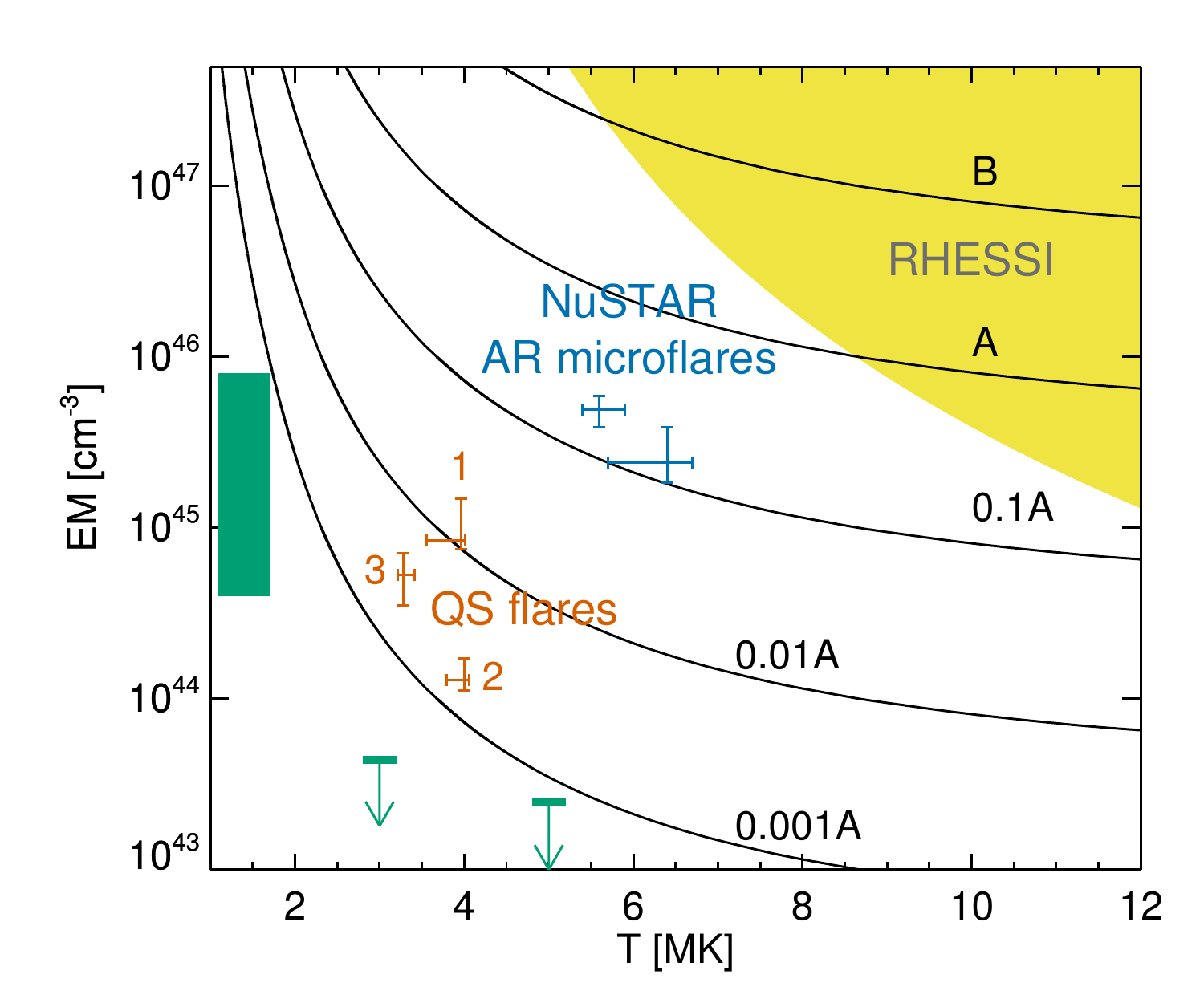}
\caption{Three analyzed events (orange) in the $T-EM$ parameter space, together with two active region microflares (blue) observed in previous NuSTAR solar campaigns. The quiet Sun network flares observed with \textit{Yohkoh}/SXT \citep{Krucker97} are depicted with the green box, together with the estimated upper limits in the temperature range of our QS events. GOES-13 classes between 0.001A and B are shown by isocurves. The part of the parameter space observable by RHESSI is shown in yellow.}
\end{figure}

\par What follows next? Solar observations with NuSTAR started in September 2014 and have been carried out sporadically every few months, depending on science questions addressed and solar conditions, giving 12 observations in total at the time of writing. Taking into account the EUV QS flare frequency distribution (Figure 1), we expect a few QS events of energies $\sim10^{26}$ ergs per hour within the NuSTAR FoV. This is roughly in agreement with our observations of 3 events in 1.5 hours of data. We overplot our observations in the frequency distribution plot in Figure 1 as a brown shaded box. The height of the box represents uncertainty in determining the number of events in the low-statistics regime following the approach of \cite{Gehrels86} and taking the conservative 99\% confidence interval, while the width of the box represents the thermal energy range of our events. 
\par As the Sun's activity decreases towards solar minimum in 2019/2020, we expect progressively better conditions for observations of QS flares. We can get an estimate of this by inspecting detector livetimes and count rates of the observed events. The data for flare 3 are taken here as an example. We emphasize the following points that will improve the sensitivity during optimal observing conditions:
\begin{enumerate}
\item Livetime could improve by a factor of $1/0.59\approx1.7$ in periods of low solar activity.
\item NuSTAR detected 900 counts above 2.5 keV during the event, with background contributing $\sim$3\% of the emission (see Figure 3). In the absence of any activity during solar minimum observations, we expect ghost-rays to largely disappear, reducing the background emission to values close to zero. The spectral analysis could then be performed with many fewer counts than we observed for flare 3; an improvement in sensitivity of up to a factor of 10 seems feasible.
\item Counts below 2.5 keV, where NuSTAR calibration is inaccurate due to threshold uncertainties and ghost-ray influence is strongest, have not been used for spectral fitting. In the absence of ghost-rays, however, using counts down to 1.6 keV can be used for flare detection. While spectral fitting will be affected by uncertainties in calibration below 2.5 keV, we might still get acceptable energy estimates. Moving the lower energy limit down to 1.6 keV would increase our statistics by a factor of 4.
\end{enumerate}

Combining these factors would lead to a sensitivity increase of a factor of $\sim70$. Assuming the same flare temperature, NuSTAR could observe QS flares with EMs of $\sim8\times10^{42}\textrm{ cm}^{-3}$ and  thermal energies of $\sim7 \times 10^{25}$ erg. Assuming the flare frequency distribution index of 2, we would expect $\sim15$ events per hour within the NuSTAR FoV. Of course, smaller events might have lower temperatures and/or different areas than the events presented here, making if difficult to estimate a lower limit of the energy content that can be reached. Even if we do not reach such low energies, observing even a few events per hour would be a significant step forward to a statistical study, which would provide further insights into the energy content and heating processes in the faintest impulsive events on the Sun.

\acknowledgments

This work made use of data from the NuSTAR mission, a project led by the California Institute of Technology, managed by the Jet Propulsion Laboratory, and funded by NASA. M.K. and S.K. acknowledge funding from the Swiss National Science Foundation (200021-140308). I.G.H. is supported by a Royal Society University Research Fellowship. LG was supported by an NSF Faculty Development Grant (AGS-1429512). We thank the referee for the thorough reading of the manuscript and the helpful comments that substantially improved the paper.

\end{document}